\documentclass[journal]{IEEEtran}
\ifCLASSINFOpdf
   \usepackage[pdftex]{graphicx}
\else
  \usepackage[dvips]{graphicx}
\fi
\usepackage{url}

\usepackage{color}

\begin{document}
%
\title{High Dynamic Range Imaging Technology}
%
%
%

\author{Alessandro~Artusi,
            Thomas~Richter,
            Touradj~Ebrahimi,
            Rafa{\l}~K.~Mantiuk,
}
%
%

\markboth{IEEE SPM Magazine,~Vol.~34, No.~5, September~2017}%
{Shell \MakeLowercase{\textit{et al.}}: Bare Demo of IEEEtran.cls for IEEE Journals}
%

\IEEEpubid{ \makebox[\columnwidth]{DOI: 10.1109/MSP.2017.2716957 \~\copyright 2017 IEEE. \hfill} \hspace{\columnsep}\makebox[\columnwidth]{ } }


\maketitle

\IEEEPARstart{I}n this lecture note, we describe high dynamic range (HDR) imaging
systems; such systems are able to represent luminances of much larger
brightness and, typically, also a larger range of colors than
conventional standard dynamic range (SDR) imaging systems. The larger luminance
range greatly improve the overall quality of visual content, making it
appears much more realistic and appealing to observers. HDR is one of
the key technologies of the future imaging pipeline, which will change
the way the digital visual content is represented and manipulated
today.



%
\IEEEpeerreviewmaketitle

\section{Prerequisites}
\label{sec:Prerequisites}

Essential knowledge of linear algebra, image/signal processing and
computer graphics is desirable. The basic aspects of High Dynamic Range (HDR) imagery
are required for the full comprehension
of this lecture note. The readers are invited to consult \cite{Artusi+2016b} for acquiring this 
basic know-how before to proceed with the reading of this lecture note.

\section{Relevance}
\label{sec:Relevance}

Due to the availability of new display and acquisition technologies,
interest in HDR increased significantly in the past years. Camera
technologies have greatly evolved providing high quality sensors that
generate images of higher precision and less noise; the market offers
now displays that are able to reproduce content with higher dynamic
range, peak luminance and color gamut.  
These advances are opening a large number of applications that
span from broadcasting to cinema, manufacturing industry to medical. 
This is also demonstrated by activities taking place nowadays within the standardization 
communities, i.e., JPEG, MPEG and SMTPE. New standards have been created for still HDR images, i.e.,
ISO/IEC 18477 JPEG XT~\cite{Artusi+2016}, others are under development. All these
activities are largely driven by industry, a strong indication that business cases
around HDR will emerge in the near future.

\section{Problem Statment and Solution}
\label{sec:ProblemStatment}
\subsection{Problem Statment}
\begin{figure}[t]
\centerline{
\includegraphics[width=1\columnwidth]{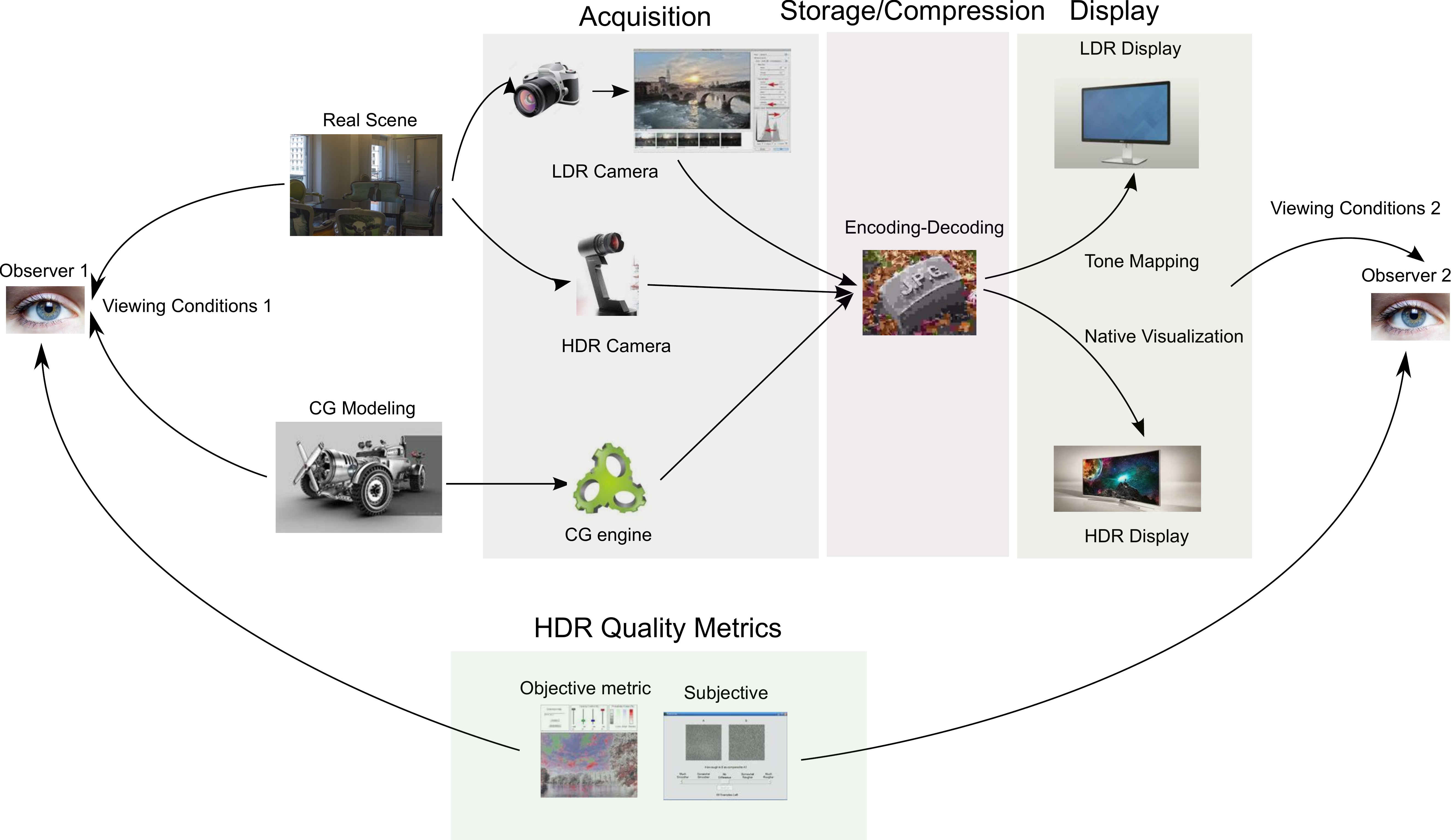}
}
\caption{HDR imaging pipeline: acquisition (greyish), storage (violet),
  display (yellowish) and evaluation (greenish).}
\label{fig:HDR-pipe}
\end{figure}

The problem to be solved consists of the development of an imaging and video system pipeline capable of representing a wider
range of luminance and colors values compared to the traditional, standard dynamic range (SDR) system pipeline. The idea
is to design a complete system, which incorporates acquisition, storage, display and evaluation subsystems, 
as shown in Figure \ref{fig:HDR-pipe}.

\subsection{Solution}

\IEEEpubidadjcol

\subsubsection{Acquisition}
\label{sec:Acquisition}
Two major ways exist to generate HDR content, either generating scene
through computer graphics tools or through the acquisition of real world
scene with a camera. Rendering pipelines for computer-generated graphics
integrate tools such as physically-based lighting simulations that use
physical-valid data of the scene and the environment, i.e., light sources and
object materials. The models used there are capable of simulating physically
plausible behavior of the light of the scene within a specific environment
and generate plausible images from an abstract scene description.

The second method acquires HDR images from real-word scenes; today, high
quality \textit{digital single-lens reflex} (DSLR) cameras are available with  sensors capable of 
capturing 12-to-16-bits per color channel. However,
many portable devices, such as mobile phones and lower quality digital
cameras are equipped with cheaper, lower performing hardware whose precision
is limited to 10 bits or even lower. 

\begin{figure}[t]
\centerline{
\includegraphics[width=1\columnwidth]{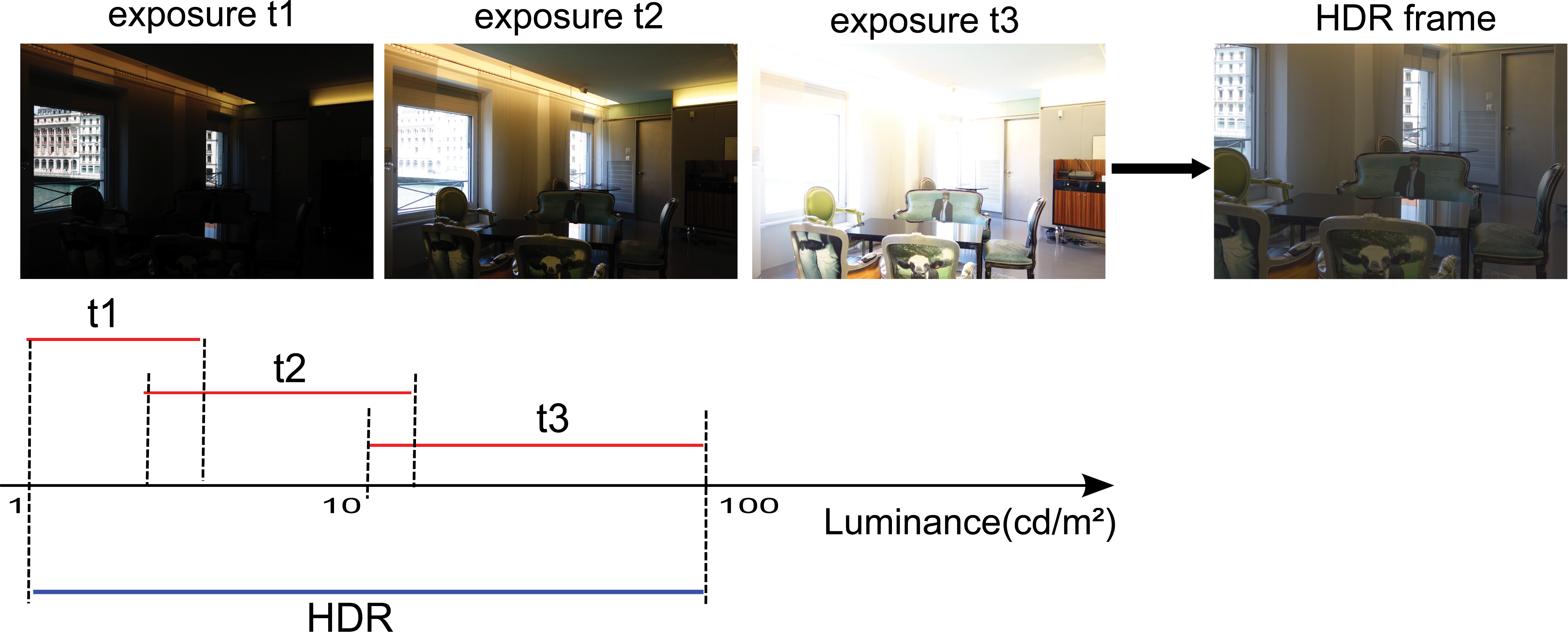}
}
\caption{Multi-exposure  approach used to capture an HDR image - (left)
  three images taken with three different exposure times $t1$, $t2$ and $t3$,
  with the different portions of the dynamic range of the scene captured by
  the exposure time $ti$ - (right) the reconstructed HDR image. The reconstructed HDR image is tone mapped for 
  display purposes.}
\label{fig:Multi-Exposure}
\end{figure}

For such a device, only a small subset of the available dynamic range of the
scene can be captured, resulting in overexposed and underexposed areas of
the acquired image. To overcome this limitation, one can capture different
portions of the dynamic range of the scene by varying the exposure time. 
The resulting images are then first registered, i.e. aligned to each
other, before a camera response function is estimated from them. This
function describes, parametrized by the exposure time, how luminances
received by the sensor are mapped to pixel values; its inverse allows
to estimate physical quantities of the scene from the acquired
images. 
Finally, a weighted
average over the pictures generates an HDR
image~\cite{Debevec+1997}. The selected weights indicate
the contribution of each frame at a given position to the final HDR sample
value. An example of multi-exposure approach is depicted in Figure
\ref{fig:Multi-Exposure}; here three images of the same scene were taken,
varying the exposure time. A (tone mapped) version of the resulting HDR is
also depicted.
A typical problem of the multi-exposure method is the misalignment of the
images, either due to movements in the scene or by the camera
itself \cite{Sen+2016}. Merging such images without further processing results in ghosting
artifacts in the HDR output. Here below, 
such defects can be classified as follows:

\begin{itemize}
\item Global misalignment due to camera motion, e.g. camera movement or
  rotation. This type of misalignment affects all pixels of the image
  causing ghost artifacts that can be removed through image registration.
  
\item Local misalignment due to moving objects in the scene, only affecting
  portions of the image. Such defects arise if the time between the
  individual exposures is larger than the typical time within which an
  object moves in the scene. For example, some objects may be occluded in
  one of the images, but are visible in others. 

\item Local and Global misalignments combining the two previous types. A
  typical example is that of a camera that follows a free path, acquiring a
  scene composed of dynamic objects.
\end{itemize}

\subsubsection{Storage and Compression}
\label{sec:Storage}
A na{\"i}ve analysis of HDR images reveals that uncompressed HDR data would
typically require four times more storage capacity than SDR data. Clearly, this view is oversimplifying the situation, but it should at least
indicate the need for a data format that is more compact. Various better alternatives exist in the field, amongst them
half-float (as used by OpenEXR), RGBE, LogLuv encoding, and representation
of sample values in a perceptually uniform color space through an \textit{electro-optical
transfer function} (EOTF). All these convert a direct, floating point
representation into a more efficient data format that requires less bits,
while still providing a higher dynamic range and more precision than an SDR
format.

If the precision of the HDR format is not sufficient, then quantization
defects such as banding will become visible.

We now discuss a selection of popular HDR formats. Half-float precision is a
compact representation for floating point values where one bit is used for
the sign, 5 bits for the exponent and 10 bits for the mantissa. 
The advantage that the half-float representation is offering is that it is
as flexible as the regular single precision floating point format at half of
the storage cost. However, since the maximum value representable by this
format is 65535, sample values should be calibrated by a common scale
factor, i.e., represented in ``relative radiance'' to be able to represent
the full dynamic range in the image.

The RGBE format takes advantage of the fact that the color components of an
RGB image are highly correlated and that they have usually very similar magnitude.
RGBE thus only stores one common scale
factor for all three components in the form of an exponent $E$, and the
individual channels are jointly scaled by $E$ as follows:
\begin{equation}
R_e = \lfloor{\frac{256R}{2^{E-128}}}\rfloor,
\end{equation}
for $G$ and $B$ the same equation applies. The $\lfloor$.$\rfloor$ denotes rounding down to the nearest integer.
E is the common exponent that is encoded together with the RGB
mantissas, resulting in a 32-bit per pixel representation.
\begin{equation}
E = \lceil{log_2(max(R,G,B))+128}\rceil,
\end{equation}
where $\lceil$.$\rceil$ denotes rounding up to the next integer.

A drawback of RGBE pixel encoding is that it cannot represent negative
samples values, i.e., colors that are outside of the triangle spanned by the primary
colors of the underlying RGB color space. A possible remedy is to code
colors in the XYZ color space taking only positive numbers by definition,
then giving rise to the XYZE encoding.

In both cases, however, errors are not uniformly distributed perceptually speaking,
a problem that is partially solved by the LogLuv encoding. There, the luminance is
coded logarithmically in one sign bit, 15 mantissa bits and another 16
bits to encode the chroma values $u_e$ and $v_e$. 

Logarithmic encoding of luminance values is a common trick used in many HDR
encodings: When the same magnitude of distortion is introduced in low and
high luminance image regions, one finds that artifacts will be more visible
in low luminance regions as human vision follows approximately a logarithmic
law --- this is also known as Weber's Law in the
literature.

However, more accurate models of human vision exist that map physical
luminance (in nits, i.e., candela per square meter) into the units related to the
\textit{just-noticeable-differences} (JNDs). Such a mapping, namely from
perceptually uniform sample space to physical luminance, is also
denoted as electro-optical transfer
function ( EOTF). Studies have shown that under such a mapping 10 to 12 bits are 
sufficient to encode luminances between $10^{-4}$ to $10^{8}$ nits without visible banding. 

HDR file formats that are making use of these HDR pixels representation have
been proposed, and the three most widely used are Radiance HDR, the Portable
File Format (PFM) and OpenEXR. Radiance HDR, indicated by the file extension
{\tt .hdr} or {\tt .pic}, is based on RGBE or XYZE encoding, plus a minimal
header. A very simple run-length coding over rows is available.

PFM is part of the ``portable any map'' format, and is indicated by the
{\tt.pfm} extension. The header indicates the number of components and a
common scale factor of all sample values; the sign of the scale factor
denotes the endianness of the encoding. The actual image pixels are encoded
as RGB triples in IEEE single precision floating point.

OpenEXR uses as file extension {\tt .exr}, and it has been developed by
Industrial Light and Magic in the 2002, along with open source
libraries. Due to its high adoption it has become the de-facto standard file
format for HDR images, especially in the cinema industry. This file format
supports three pixel encoding formats: half-float (16-bit float), 32-bit
float and 32-bit integer. It also includes various lossy and lossless image
compression algorithms.

We recently see the adoption of HDR technologies into products such as
cameras with improved sensors, displays providing higher dynamic range
and/or larger color gamut. Unfortunately, interoperability at device level
is still at its infancy, making it difficult to exchange images between
various devices, or various vendors which try to lock-in customers through
proprietary formats~\cite{Artusi+2016}.

As for images, two international standards are already available that support HDR content, namely
ISO/IEC 15444, ITU-T T.800 JPEG~2000 and ISO/IEC 29199, ITU-T T.832 JPEG~XR. 
Despite the fact that they support lossless
compression, their limited adoption by the market may be correlated with
their lack of backward compatibility with existing JPEG
ecosystem~\cite{Artusi+2015}. Industry players are typically reluctant to change technology in their production pipeline to cope 
with adoption of newly established standards.
A migration path from existing to new solutions,
allowing a gradual transition from old to new technology helps them to keep
the investments low.

\begin{figure}[b]
\centering
\includegraphics[width=\linewidth]{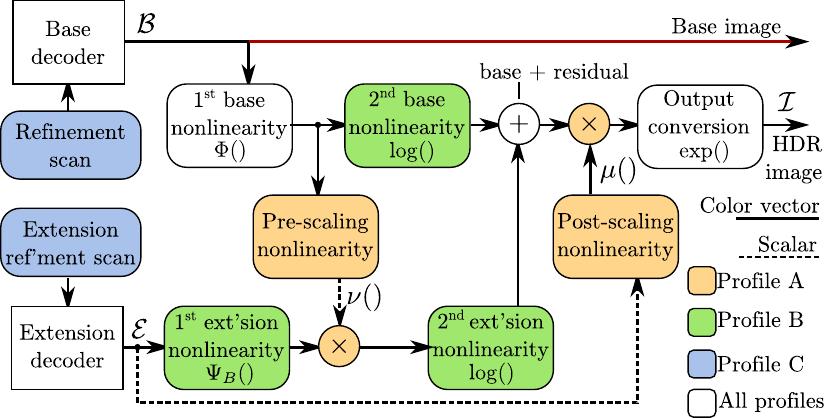}
\caption{\small{The simplified decoding workflow for JPEG XT
    standard\cite{Artusi+2015}. ${\cal B}$ is the base layer and is always
    represented as a JPEG codestream with 8-bit per sample.  ${\cal E}$ is
    the extension layer that used in conjunction with ${\cal B}$ allows the
    reconstruction of the HDR image.}}
\label{fig:JPEGXT}
\end{figure}

To address this issue, the Joint Photographic Experts Group (JPEG) formally
known as ISO/IEC JTC1/SC29/WG1, began in the 2012 the standardization of a
new standard technology called ISO/IEC 18477 JPEG XT \cite{Artusi+2016}. The JPEG~XT
image coding system is currently organized into nine parts that define the baseline
coding architecture (the legacy JPEG codestream 8-bit mode), an extensible
file format specifying a common syntax for extending the legacy JPEG, and
application of this syntax for coding integer or floating point samples
between 8 and 16 bits precision~\cite{Artusi+2016}.

This coding architecture is then further refined to enable lossless and
near-lossless coding, and is complemented by an extension for representing
alpha-channels \cite{Artusi+2016}. Due to its flexible layered structure, the
JPEG~XT capabilities can be extended into novel applications such as
omnidirectional photography, animated images, structural editing as well as
privacy and security that are under examination and development \cite{Richter+2016}.

In practice, JPEG~XT can be seen as a superset of the 8-bit mode JPEG where
existing JPEG technology is re-used whenever possible; this, in particular,
allows to encode an HDR image purely on the basis of legacy JPEG
implementations. JPEG~XT is a two layered design, of which the first layer
represents the SDR image. It is encoded in JPEG, with 8-bits per sample in
the ITU BT.601-7 RGB color space (base layer ${\cal B}$), see Figure \ref{fig:JPEGXT}.
The extension layer ${\cal E}$ includes the additional information to reconstruct
the HDR image starting from the base layer ${\cal B}$.

Concerning video compression, some recent standards are
providing options to encode video in high bit precision, i.e., up to 12 bits
for ISO/IEC 14496-2 and ISO/IEC 14496-10 AVC/H.264. These modes are defined
in the profile Fidelity Range Extensions (FRExt), and for ISO/IEC 23008-2 ITU-T-H.265 (HEVC)
in the Format Range Extension (RExt).

The H.264/AVC extensions build upon an EOTF that covers a dynamic range of up to $2.5$ magnitudes; while
sufficient for consumer applications, this is a limitation for typical HDR
content.

H.265/HEVC recently integrated a transfer function for HDR video content
that pre-quantizes data to a 10 or 12 bit domain which is then taken as
input by the HEVC encoder. This EOTF, denoted as ST2084 --- Hybrid
Log-Gamma --- is designed for luminances up to $10,000$ nits.
Finally, guidelines on how to encode HDR video content with HEVC, have been provided
in ISO/IEC 23008-14 and 15.

Similar to JPEG~XT, a backward compatible solution for HDR video encoding has been presented by 
Mantiuk et al.\cite{Mantiuk2006a}. Recently a signaling mechanism to support backward compatibility, 
has been integrated into the HEVC standard (ISO/IEC NP TR 23008-15). 
The backward-compatibility is achieved as in the case of HDR still image
encoding described above. A base layer encodes the SDR frames, and an
extension layer hidden from the base includes the necessary information to
extend the dynamic range. To improve encoding
performance, the redundancy information is minimized through the
decorrelation between the SDR and HDR streams, achieving a reduction in size
of the HDR stream to about $30\%$ of the size of the SDR stream. Invisible
noise reduction is also used to remove details that cannot be seen in the
residual stream prior to encoding.

\subsubsection{Display}
\label{sec:Visualization}
The native visualization of HDR content is limited by the
physics of the display. Despite the fact that the current technology on the
market can guarantee high contrast ratio, this is achieved by lowering the
black level. However, the peak luminance remains limited, restricting the
available dynamic range for bright images. Even with enhanced contrast, many
display panels offer only a limited precision of 8 or at most 10 bits per
color channel, and not all of them support a wide color gamut neither.

\begin{figure}[t]
\centerline{
\includegraphics[width=1\columnwidth]{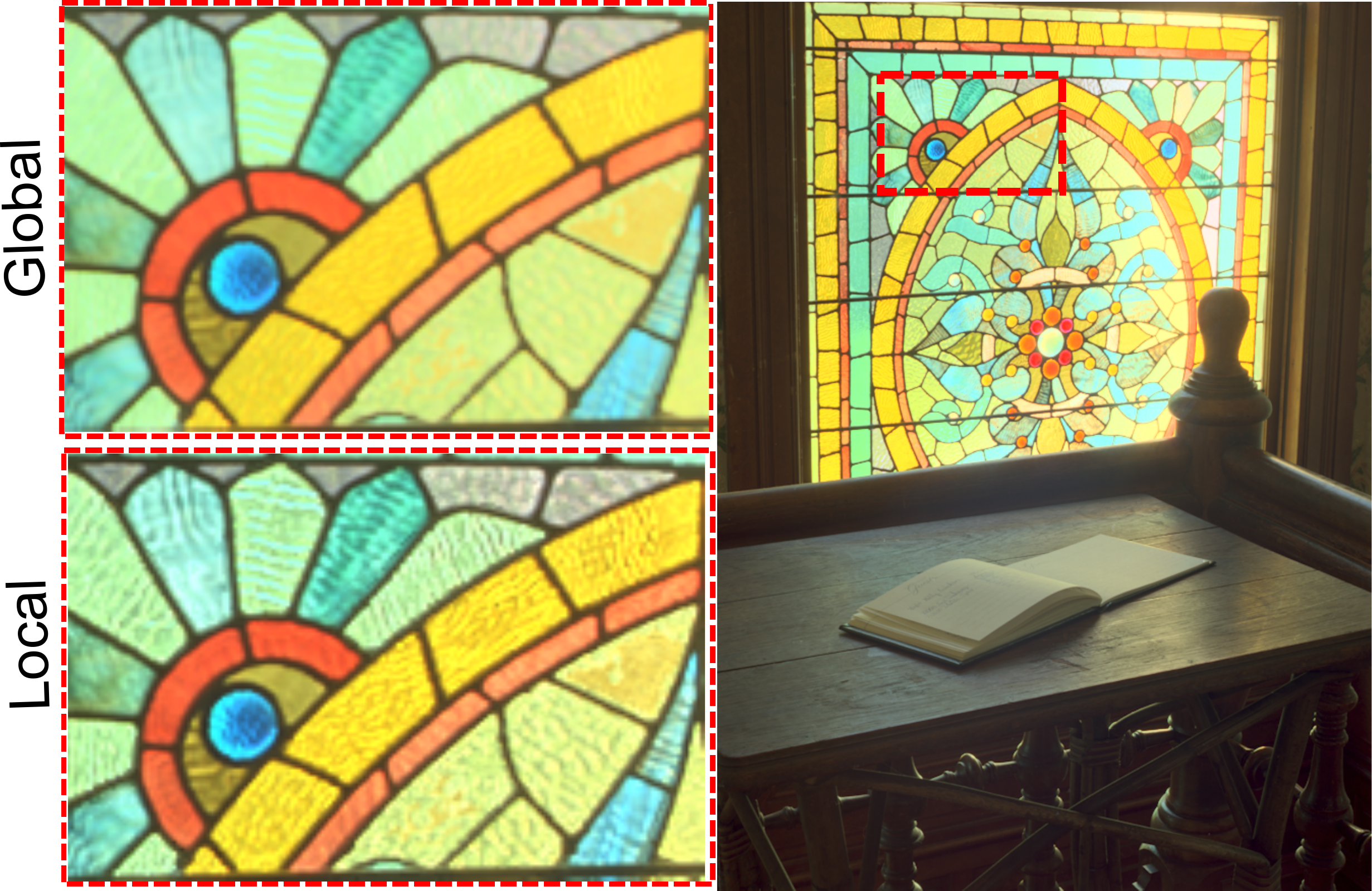}
}
\caption{Global vs Local - Global approach results in lost of details in high
  contrast regions. The HDR frame is tone mapped for display purposes.} 
\label{fig:GvsL}
\end{figure}

Tone mapping is a process that compresses the dynamic range of an input
signal to that available by the display or the printing process while
keeping the visualization convincing. 
Tone mappers can be roughly classified into global and local approaches. The
former is applying the same tone curve on the all image pixels.  The latter
takes the spatial position and its surrounding into account; by that, local
operators can take advantage of known effects of the human visual system
such as local eye adaption to the luminance. While the former is simple and
efficient, it may fail to reproduce details in high contrast image regions
(see Figure \ref{fig:GvsL} (top-left)). Although the latter can reproduce details
in such regions better, see Figure \ref{fig:GvsL} (bottom-left), it often comes at the cost of increased complexity
and computational time; it may also introduce artifacts around edges. 

Despite this classification, we may also categorize the tone mappers based on their intent. Three main
categories of tone mappers can be identified: 
based on visual system, for scene reproduction and for best subjective quality.
The first aims at integrating into the tone mapper mechanisms that simulate various
aspects of the human visual system. This includes glare, luminance and chromatic adaptations, night vision, etc.
The second category attempts to reproduce the best match in color gamut 
and dynamic range available for the display on which the image will be visualized. This is achieved
through the preservation of the original scene's appearance.
The last category produces images with most preferred subjective quality. Typical examples are
operators with parameters that can be adjusted to achieve a
specific artistic goal.

\begin{figure}[htb]
  \center{
\includegraphics[width=0.45\textwidth]{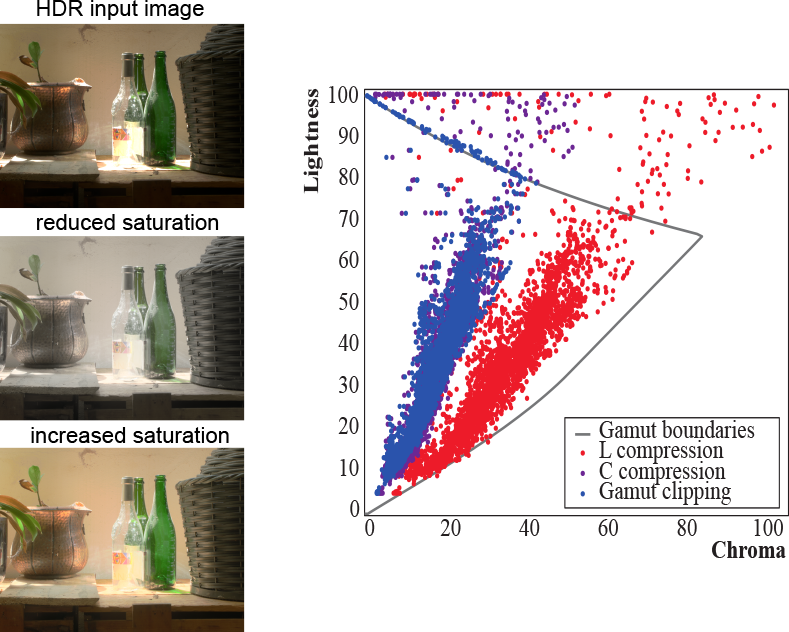}
   }
  \caption{Tone mapper drawbacks - (left) changes in appearance due to
    either a reduction or an excessive saturation, - (right) pixels may be within the destination
    gamut only for the lightness channel $L^*$; however, their chroma
    channel may still be out of gamut. Here the HDR input image has been tone mapped for display purposes. 
    Left image courtesy of Francesco Banterle and right image courtesy of Tania Pouli.}
  \label{fig:Saturation}
\end{figure}

\textbf{Color Correction}
Dynamic range mismatches between the HDR data and display devices, as previously shown, are typically
handled by tone mappers, focusing on one dimension of the color gamut, along the luminance direction.
This generates two major drawbacks. First, appearance effects are often
ignored, leading to images which may appear poorly or too saturated as shown in Figure
\ref{fig:Saturation} (left) \cite{Pouli+2013}. Second, such a tone mapper may
not guarantee that all the sample values of the tone mapped image are within
the available target, as shown in Figure \ref{fig:Saturation} (right). Even though the
output luminance may be reproducible by the display, the chrominance may
fall out of the available gamut, resulting in clipping of extreme
colors. This clipping may again introduce hue shifts and image
defects.~\cite{Sikudova+2016}.

To improve the saturation of the tone mapped image,  a simple solution is to introduce an adjustable parameter which 
allows to control the overall saturation  of the tone mapped image~\cite{Mantiuk+2009eg}. In the following, let $p$ be a parameter in
$[0,1]$, then
\begin{equation}
 I_t = \left(\frac{I_o}{L_o}\right)^p L_t,
 \end{equation}
Here $I_o$ is the input HDR image, $I_t$ is the final output tone mapped image (both in RGB values), $L_o$ is the 
luminance of the original HDR image and $L_t$ is the
luminance of the tone mapped image. The parameter $p$ then needs to be
selected for the best --- most pleasing --- result. 
Unfortunately, the simple solution presented above does not only
adjust the saturation, it also implies a luminance shift. Controlling $p$ to get the desired effect may be hard.
This problem can be overcome by a more careful choice of the input and
output scaling operation~\cite{Mantiuk+2009eg}:
\begin{equation}
I_t=\left(\left(\frac{I_o}{L_o}-1\right)p+1\right)L_t.
\end{equation}
While this allows better control of the luminance shift, it may cause
undesirable hue artifacts~\cite{Pouli+2013} if applied separately to each
component of an RGB image.  The value of $p$ in the above equations can
be automated based on the slope of the tone curve at each luminance
level~\cite{Mantiuk+2009eg}.
To reduce hue and lightness shifts, one may work with perceptual uniform color space to separate the color
appearance parameters such as saturation from hue and lightness. This will allow to modify the saturation, of the 
tone mapped image, to match the saturation of the input HDR image while hue and lightness of the tone mapped image 
$I_t$ will remain untouched~\cite{Pouli+2013}.
Other approaches exploit the use of color appearance models,
and extend the concept of gamut mapping of the HDR content~\cite{Sikudova+2016}.
The former approach guarantees the matching of the color appearance
attributes between the input HDR and the tone mapped images. The latter
ensures that all the tone mapped pixels are within the color gamut of the
display, minimizing the hue and luminance distortion.

\textbf{Inverse Tone Mapping}
The latest standardization trends and technological improvements push the
display features towards ultra HD, higher dynamic range (HDR), i.e., up to
$1,000$ and $6,000$ nits, and wide color gamut (ITU-R Rec. 2020).
Since traditional LCD panels with constant backlight illumination are not
able to reproduce the necessary dynamic range, HDR displays make use of a
modulated backlight. In such a display, a front-layer LCD
panel includes the color filters and provides the necessary level of details
for accurate image reproduction and a lower resolution matrix of
independently controlled LEDs modulate its illumination at a coarser level,
providing a much larger dynamic range. Optical layers and reflectors around each
LED maximize the brightness in its corresponding area of the front
LCD panel and minimize the light leakage into adjacent cells.  
Due to the coarser resolution of the back panel image quality degradation may appear, which can be
reduced through the use of post processing filtering of the displayed image.

The widespread availability of SDR content and the recent availability of displays
with larger dynamic range also made it attractive to process such content
for presentations on HDR displays. This process can be seen as the opposite
problem of tone mapping, and is thus called ``inverse tone mapping''.
The ability to reconstruct the mapping between the
pixel values encoded in the SDR image and the scene luminance values, also
known as inverse camera response function, is the desirable goal.
While it is an easy task to reconstruct the camera
response function from a series of different exposures of the same SDR
content, it is an ill-pose problem to reconstruct such an inverse when only
a single exposure of an unknown camera is available.

The camera response function models the complete pipeline from light
acquisition to SDR pixel values, including the (non-linear) sensor
response, exposure, camera post-processing (e.g. flare-removal) and
tone mapping of raw pixel values to SDR sample values. Recovering the dynamic
range for an SDR content will consist of two basic steps. First, estimate
an inverse camera response function to linearize the SDR content signal,
then adjust the dynamic range of the SDR pixel to fit it to the dynamic range
of the HDR display.
However, SDR images are presenting two major issues when expanding them
to larger dynamic range. First, the limited pixels precision, i.e., quantization to
256 values per channel, causes loss of detail and posterization. These artifacts while barely
visible in the SDR domain, can be emphasized during the expansion of the dynamic range.
Second, under and over-exposed regions in the SDR image contain very
limited information. This may lead, during the dynamic range
expansion, to regions that have the same appearance as in the original SDR image.

To solve the first problem, advanced filtering is needed before
boosting the dynamic range of the SDR image. Bilateral filtering is an
example: by tuning its parameters properly, high- and low frequencies can be
processed separately avoiding some of the typical artifacts of
range-expansion. 
While lost image content cannot be recovered in any way, to solve the second problem, inpainting may at
least generate plausible image details in under- or overexposed image
regions, provided the regions are sufficiently small and enough details
are available around them.

\subsubsection{HDR Quality Indices}
\label{sec:HDR Quality Indices}
The evaluation of the quality of an image or video 
is one of the fundamental steps in understanding whether the algorithm is
capable of achieving a level of quality acceptable for a specific
application. Depending on whether the original source is available when
assessing a somewhat distorted image or video, one distinguishes between
``full reference'' and ``no reference'' quality indices. If only partial
information on the original is available, they are called ``reduced
reference'' indices. In a second dimension, we can distinguish between ``objective'' and
``subjective'' quality indices. In the former method, a computer algorithm quantifies the differences
between a reference and a test image or video. Such an algorithm may include
a model of the human visual system, and then evaluates the visibility of image
defects in terms of its observer's model. The latter method, evaluates quality through studies by human
observers. Based on a particular test methodology, observers are asked to
qualify characteristics of single or pairs of visual stimuli in form of image or video
and to provide a score on a scale, or a relative rating between multiple presentations. Certainly, the second method is capable of catching all aspects of human
vision and is thus more appropriate to evaluate (or even define) the quality
of an image or a video. It is, however, also very resource and time
consuming and only a limited number of media artifacts can be rated by such
a method. Objective quality indices, as computer implementations, are more
convenient as they allow automatic assessment. However, they are less reliable
in estimating the overall image quality as their assessment is based on a limited mathematical model.

While reliable objective full-reference metrics are known and have been
studied multiple times, no-reference quality prediction by computer
algorithms is a much harder problem. Subjectively, both full and
no-reference methods are in use, though might answer slightly different
questions. Full reference methods measure ``fidelity'' --- how close is the
distorted image to the reference --- while no-reference methods rate the
overall ``quality'' of a presentation.

In the following, we will focus on full-reference objective quality
indices. Here one can again distinguish between display-referred and
luminance-independent metrics. The former expect that the values in images or video correspond to the
absolute luminance emitted from a display on which a presentation is
shown. The latter accept any relative radiance values as input. They assume that
human vision is approximately scale-independent, a property that is
equivalent to the 
``Weber's Law''.
Generally, the objective metrics designed for SDR such as PSNR and SSIM
are ill-suited for HDR content. These metrics take as input a gamma corrected image and consider this content 
in an approximate perceptually uniform space. 
However, this assumption is valid for CRT displays that are working typically in low luminance range (0.1
to 80 nits). This is not anymore valid for brighter displays. Here distortions that are barely visible in CRT displays, will be noticeable. 
A simple encoding of the physical luminance that makes objective metrics for SDR content applicable for HDR content is to transform
 luminance values into a perceptually uniform (PU) space. 
Two major objective metrics for evaluating HDR content directly have emerged
lately \cite{Narwaria, Aydin08}. Both are full-reference human visual system based metrics.
HDR-VDP-2 is an objective metric capable of detecting differences in
achromatic images. For that, it includes a contrast sensitivity model
inspired by the properties of the human visual system for a wide range of
luminance values. The metric takes test and reference HDR images as input
which are then mapped first to a perceptual space and frequency-filtered in
several orientation and frequency specific subbands modeling the first
layer of the visual cortex. In each subband, a masking model is applied. The
subband wise difference of the cortex-filtered output is then used to
predict both visibility as the probability of defect-detection, and quality,
as the perceived magnitude of distortion. 

The \textit{dynamic range independent} (DRI) metrics attempt to evaluate image quality
independent of the dynamic range of the two images to be compared.
 If the dynamic range would be identical, the pixel-wise difference between test and
reference images would already provide an indicator of the visible artifact
to be measured. If the dynamic range is different, though, a per-pixel difference could either be
due to an image defect degrading image quality, or due to the change of the dynamic range. In the latter, 
the visible differences in the test image should not be classified as visual artifacts. To distinguish between the two causes, such metrics apply a model of the HVS
based on the detection and classification of visible changes in the image
structure. These structural changes are a measure of contrast, and can be
categorized as follows\cite{Aydin08}: 
\begin{itemize}

\item loss of visible contrast, i.e., if a contrast is visible in the
  reference image but is not anymore visible in the test image. This happens,
  for example, if the tone mapper compresses details so much such that they
  become invisible after tone mapping.
  
\item amplification of invisible contrast, i.e., the opposite of the above
  effect. This type of degradation is typical for inverse tone mapping when,
  due to contrast stretching, contouring artifacts start to appear.

\item reversal of visible contrast, i.e., if the contrast in the test image
  is the inverse of the contrast in the reference image. Such defects
  appear, for example, due to clipping after tone mapping.
\end{itemize}

The evaluation of HDR video content is also a very important issue in
various applications and standardization activities. Recently, the HDR-VQM
metric has been proposed~\cite{Narwaria} to provide a feasible
objective metric to evaluate quality in HDR video content. Video quality is computed
based on a spatio-temporal analysis that relates to human eye fixation
behavior during video viewing.

\section{What we have learnt}
\label{sec:conclusion}

Based on this article, readers could have learned what HDR imagery is, including all steps involved in its specific imaging pipeline, and what extra features it is capable to provide to the end-user. 
In particular, HDR imagery is conveying to the end-user an extraordinary experience, when compared to 
the traditional digital imaging as known today, e.g., 8-10 bits. To better understand improvements introduced by HDR content, 
which is perceived by the end-user, one can compare it to what happened about 50 years ago when television moved from black/white to color.


\section{Acknowledgments}
EPFL author acknowledge the Swiss State Secretariat for Education, Research and Innovation (SERI), for the European 
Union's Horizon 2020 research and innovation programme under the Marie Sklodowska-Curie grant agreement
No 643072 - QoE-Net.


%








\bibliographystyle{IEEEtran}
%
\bibliography{IEEEpaperHDR}

%

\begin{IEEEbiographynophoto}{Alessandro Artusi}
(artusialessandro4@gmail.com)
is a researcher at KIOS Research and Innovation 
Center of the University of Cyprus. He is a
member of the ISO/IEC/JCTC1/SC29/
WG1 Committee (also known as JPEG),
one of the editors of the JPEG XT standard, and
recipient of the Emerging Standards
Awards from the British Standard
Institute. 
\end{IEEEbiographynophoto}

\begin{IEEEbiographynophoto}{Thomas Richter}
(richter@tik.uni-stuttgart.de)
is a researcher at the TIK Computing Center of
the University of Stuttgart. Member of SC29WG1 since 2003.
\end{IEEEbiographynophoto}

\begin{IEEEbiographynophoto}{Touradj Ebrahimi}
(Touradj.Ebrahimi@epfl.ch) 
is a professor at
Ecole Polytechnique F\'{e}d\'{e}rale de
Lausanne heading its Multimedia Signal
Processing group. He is also the convenor
(chair) of the JPEG Standardization
Committee.
\end{IEEEbiographynophoto}

\begin{IEEEbiographynophoto}{Rafa{\l} K. Mantiuk}
( rkm38@cam.ac.uk) 
is a senior lecturer at the Computer
Laboratory, University of Cambridge,
United Kingdom. He is the author of a
popular HDR image quality metric,
HDR-VDP-2; the coauthor of pfstools,
software for high-dynamic range image processing. 
\end{IEEEbiographynophoto}




\end{document}